# Band-Sweeping M-ary PSK (BS-M-PSK) Modulation and Transceiver Design

Ahmed M. Alaa

Department of Electronics and Communications Engineering Faculty of Engineering, Cairo University

Emails: aalaa@eng.cu.edu.eg

*Abstract*— Channel Estimation is a major problem encountered by receiver designers for wireless communications systems. The fading channels encountered by the system are usually time variant for a mobile receiver. Besides, the frequency response of the channel is frequency selective for urban environments where the delay spread is quite large compared to the symbol duration. Estimating the channel is essential for equalizing the received data and removing the Inter-Symbol Interference (ISI) resulting from the dispersive channel. Hence, conventional transceivers insert pilot symbols of known values and detect the changes in it in order to deduce the channel response. Because these pilots carry no information, the throughput of the system is reduced. A Novel modulation scheme is presented in this work. The technique depends on using a carrier signal that has no fixed frequency, the carrier tone sweeps the band dedicated for transmission and detects the transfer function gain within the band. A carrier signal that is Frequency Modulated (FM) by a periodic ramp signal becomes Amplitude Modulated (AM) by the channel transfer function, and thus, the receiver obtains an estimate for the channel response without using pilots that decrease the system's throughput or data rate. The carrier signal itself acts as a dynamic frequency domain pilot. The technique only works for constant energy systems, and thus it is applied to PSK transceivers. Mathematical formulation, transceiver design and performance analysis of the proposed modulation technique are presented.

## I. INTRODUCTION

In a typical wireless channel, the transmitted signal is applied to a time-variant frequency-selective channel. This means that during a specific period (*Coherence time*), the channel has a transfer function that is approximated to be constant over a specific band (*Coherence bandwidth*). These effects, namely *time* and *frequency dispersion* arise from the user's mobility and the large number of multipath reflections that represents a large delay spread and thus, susceptibility to Inter-Symbol Interference (ISI). Fading effects causes extreme degradation in the Bit Error Rate (BER) performance of wireless systems. In order to mitigate the dispersive channel effects, *Equalization* must be applied to the received signal at the receiver end. Many Equalization techniques can be used to combat ISI, those include []: Zero-Forcing (ZF), Maximum Likelihood Sequence Estimators (MLSE), Decision-Feedback Equalizers (DFE) and Minimum Mean Square Error Equalizers (MMSE). As the equalizer removes the channel effects, it is logical that the equalizer would require knowledge of the estimated channel response. For this reason, Channel estimation is mandatory for any wireless mobile receivers. Many channel estimation algorithms are proposed in literature. Most of these algorithms rely on transmitting some known data sequences by the transmitter and detecting the changes occurring in it at the receiver and thus deducing the channel response. Pilots may be *time domain pilots* (separated data bursts transmitted before every frame called a *training sequence*) or *frequency domain pilots* (usually in multicarrier systems such as OFDM). In both cases, we allocated a portion of the transmission bandwidth or a period of time to transmit those pilots, which decreases throughput and transmission efficiency. Some other techniques rely on blind equalization, where no pilots are transmitted with the signal. However, they still need a robust channel estimate in order to apply channel inversion to the received signal.

In this work, we present a novel modulation methodology that uses a new class of RF carriers. The carrier used has no fixed instantaneous frequency and instead sweeps the transmission band on periodic basis. The period for channel sweeping depends on the channel coherence time. The resultant of this carrier sweeping is that the carrier becomes AM modulated by the channel's transfer function. Channel estimation can be obtained by envelope detection of the received RF carrier. The generation of this sweeping carrier is achieved by FM modulation of a sinusoidal signal with a periodic ramp signal. Because the channel is estimated via envelope detection of the received carrier, the transmitted symbols must have constant energy and thus Phase Shift Keying (PSK) is suitable to modulate the FM sweeping carrier. We'll call this technique: Band-Sweeping M-ary PSK (BS-M-PSK) modulation. The paper is divided as follows: section II presents a system model and a mathematical model for the proposed modulation technique, section III presents the channel estimation algorithm and receiver architecture of a BS-M-PSK system. It is shown that a two-tap Rayleigh fading channel can be estimated in a simple method by processing the received band sweeping carriers. Analytical spectral analysis for the system is presented.

## II. BS-M-PSK TRANSMITTER MODEL

The BS-M-PSK modulator relies on an FM modulated carrier. The modulating signal is a periodic ramp function (sawtooth) with a period equal to the coherence time $T_c$. The periodic ramp function is defined as:

$$\emptyset(t) = Kt, \; n\frac{T_c}{2} \leq t \leq n\frac{T_c}{2} + T_c \quad (1)$$

where $K$ is an arbitrary constant. This signal is then use to modulate the carrier signal of carrier frequency $f_c$ via Frequency Modulation. This allows the instantaneous frequency of the carrier to be changed from the minimum frequency to the maximum in the transmission bandwidth.

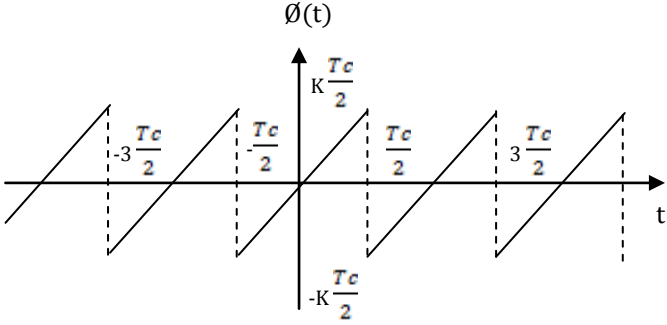

Figure 1 the periodic ramp signal (saw tooth) with a period equal to the coherence time, this signal is used to modulate the carrier using FM modulation

By applying FM modulation to the sinusoidal carrier, the instantaneous frequency of the carrier at the start of the period is shifted back from the central carrier frequency by an amount decided by the *frequency deviation*. This instantaneous frequency keeps increasing by a rate decided by the ramp's slope till it reaches zero amplitude that corresponds to the exact central frequency. The ramp signal then has positive values which reflect on the carrier tone shifted to frequencies higher than the central carrier frequency. Figure 2 shows the carrier signal after being FM modulated by the saw tooth signal.

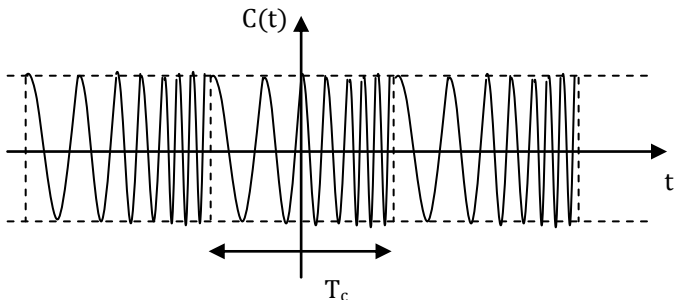

Figure 2 the sinusoidal carrier after being FM modulated by saw tooth signal

For a frequency modulation constant $K_f$, then:

$$C(t) = \cos\left(2\pi f_c t + K_f \int_{-\infty}^{t} \emptyset(\lambda) d\lambda\right) \quad (2)$$

Assume that this RF carrier is transmitted through a fading channel, we will show that the received signal will be AM modulated by the channel's transfer function. We'll start by an intuitive analysis for the signal $C(t)$ passed by a fading channel and then, analytical derivations for the transmission of the FM modulated carrier through a fading channel will be presented. Because the signal is periodic, with a fundamental period of 1/Tc, we expect a discrete frequency spectrum (obtained by applying Fourier series representation). Assume that the multipath channel is $h(\tau)$ and the modulated carrier is decomposed into an infinite number of frequency components.

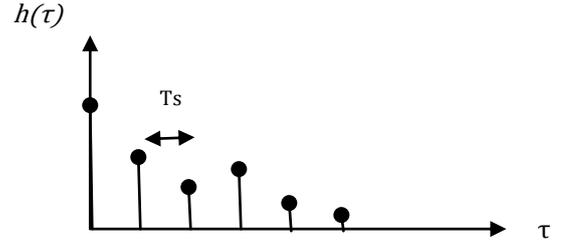

Figure 3 an example for a wideband channel impulse response (Power Delay Profile PDP)

This impulse response has a corresponding transfer function in the frequency domain, because the impulse response shown in figure 3 is composed of a set of impulses, the transfer function representing the channel will be periodic and thus, have an infinite bandwidth.

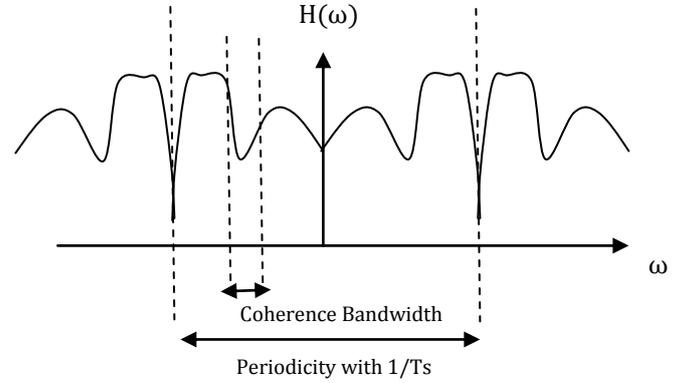

Figure 4 the wideband transfer function of the fading channel at a certain time instance

By decomposing the modulated carrier C(t) into a set of sinusoidal signals, the gain of a certain frequency component is high if the shifted and scaled versions of the sinusoid (by the shifts and scales dictated by the channel impulse response) add constructively and the gain is low if they add destructively. Because the signal C(t) has a different instantaneous frequency for every time instance, the instantaneous phase and frequency of C(t) at a time *t* are:

$$\Theta(t) = 2\pi f_c t + K_f \int_{-\infty}^{t} \emptyset(\lambda) d\lambda$$

$$\frac{d\Theta(t)}{dt} = 2\pi f_c + K_f K t, \; -T_c/2 \leq t \leq T_c/2 \quad (3)$$

The signal C(t) can be composed from an infinite set of sinusoidal signals that have different instantaneous frequencies. Assume that the period of the sawtooth signal Tc is divided into a set of *N* periods where the carrier has a constant instantaneous frequency. Thus, C(t) can be composed by multiplying the sinusoidal of a constant frequency with a pulse train with a period of $T_c$ and pulse width of $T_c/N$. The N sinusoidal signals with N distinct frequencies are multiplied by N pulse train; each is shifted from the other by $T_c/N$. By summing the resultant of the N multiplications, we obtain a signal C(t) that is characterized by a monotonically increasing instantaneous frequency. The exact depiction of C(t) is obtained when the duration of the constant frequency sinusoid tends to zero or in other words, $T_c/N = 0$, this is when $N \rightarrow \infty$ and the signal has infinite frequency components.

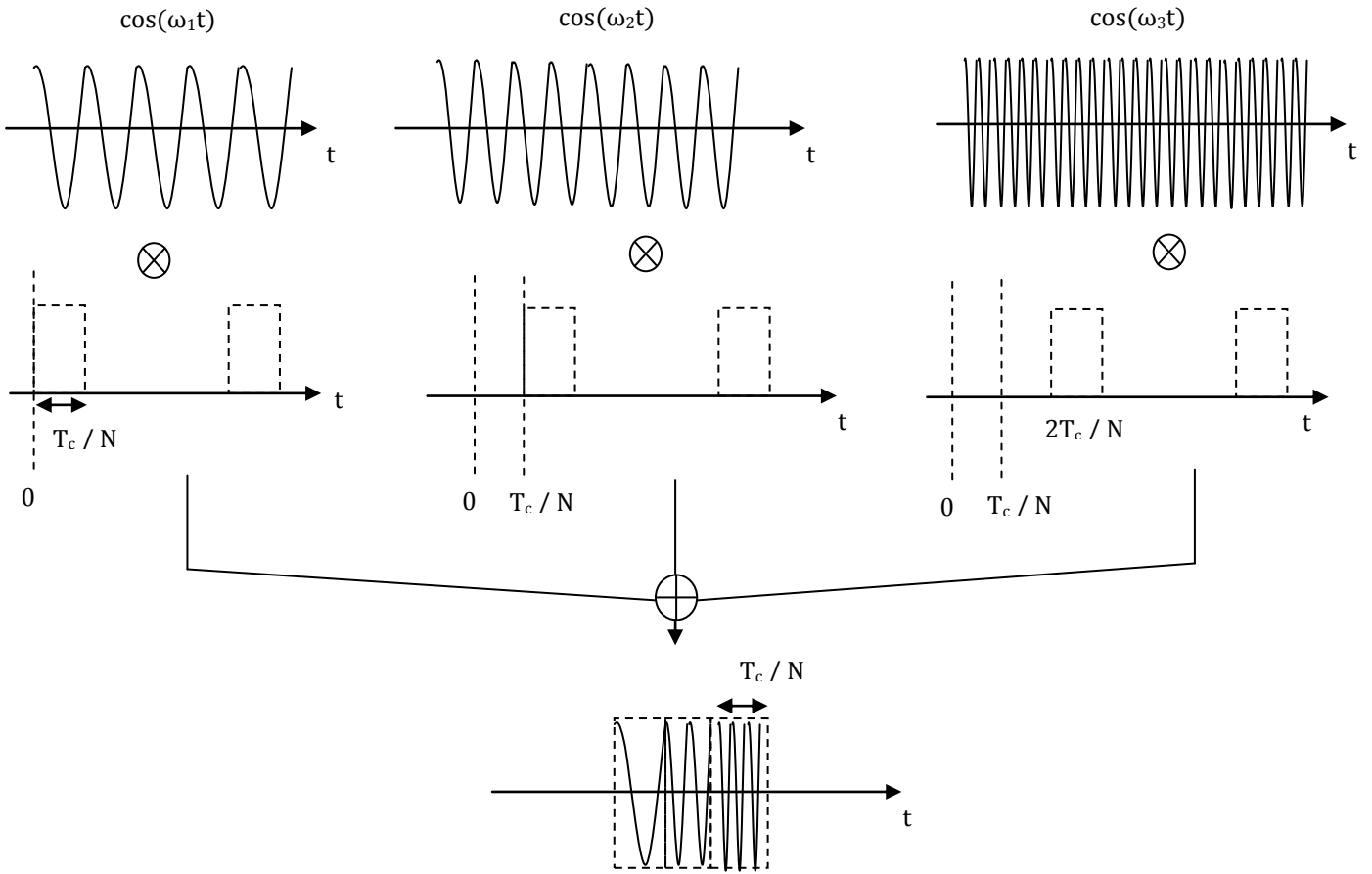

Figure 5 the construction of the FM modulated carrier signal, if the carrier sweeps from $f_{min}$ to $f_{max}$, the period is segmented into three sub-periods during which the sinusoidal has a constant frequency. Thus, the pulse train tends to be an impulse train. Every time instance has amplitude sampled from a different sinusoid with a different frequency.

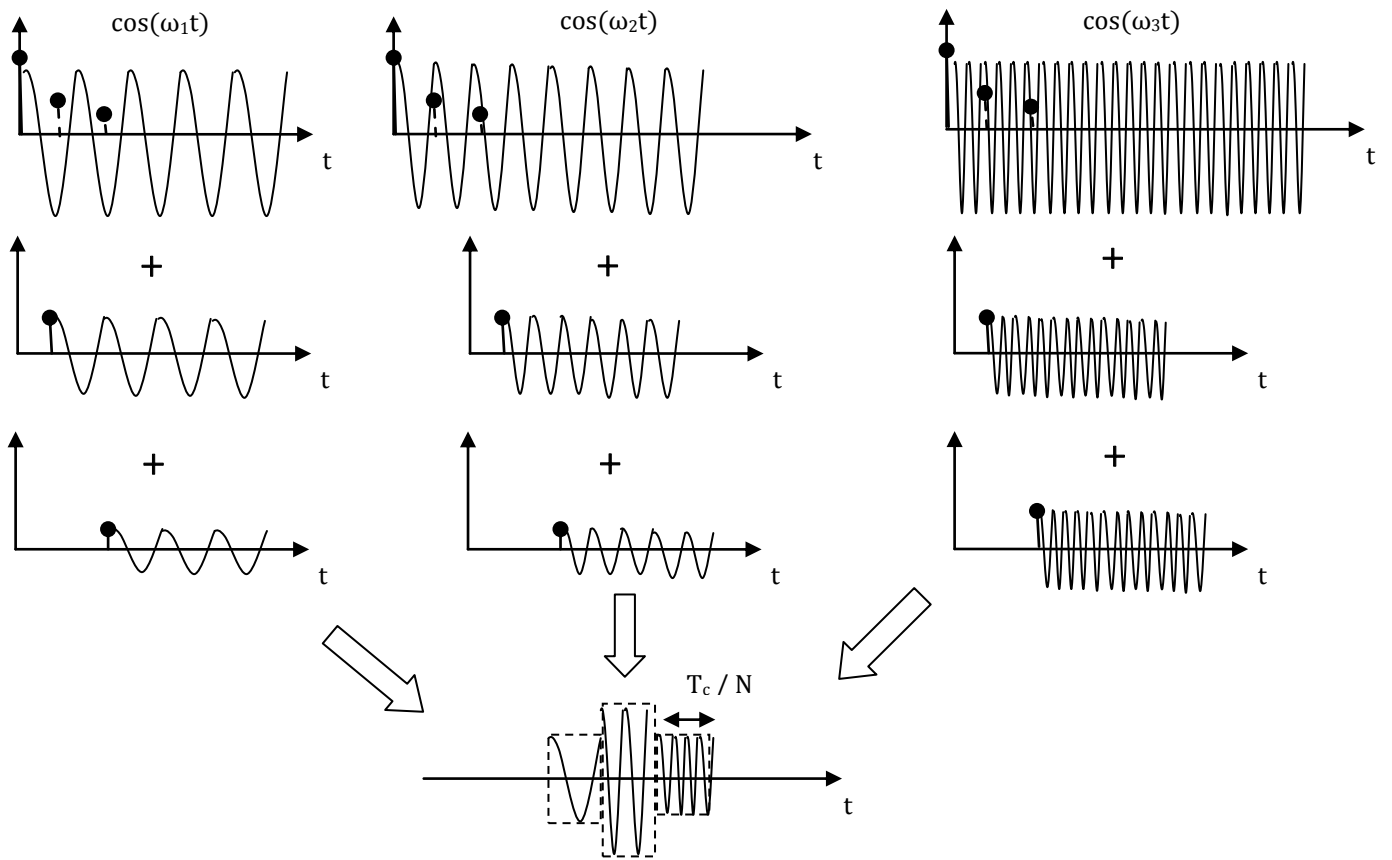

Figure 6 the result of the modulated carrier signal to a multipath channel. Different scaled and shifted versions of the signal add either constructively or destructively according to the frequency of a certain sinusoid. The resulting signal has an envelope that is equivalent to a staircase approximation of the channel's transfer function. When the period of constant instantaneous frequency tends to zero, the resulting signal has an envelope that is equal to the transfer function of the frequency selective fading channel.

The effect of the multipath channel is either constructive or destructive interference per frequency component. Every sinusoid with certain frequency will be shifted and scaled according to the multipath channel profile. According to the frequency sinusoid, the resultant of the addition is either constructive or destructive, and this defines the transfer function gain for a certain frequency bin. If we used this modulated carrier for transmission of constant energy constellation points, then the received signal's amplitude would describe the channel's transfer function. The constellation with constant energy symbols is a PSK constellation. The *orthonormal basis functions* of a BS-M-PSK system are:

$$\emptyset_I(t) = \sqrt{\frac{2}{T_s}} \cos(2\pi f_c t + K_f \int_{-\infty}^{t} \emptyset(\lambda) d\lambda), 0 \leq t \leq T_s$$

$$\emptyset_q(t) = \sqrt{\frac{2}{T_s}} \sin(2\pi f_c t + K_f \int_{-\infty}^{t} \emptyset(\lambda) d\lambda), 0 \leq t \leq T_s \quad (4)$$

It is mandatory to show that the basis functions are orthonormal; which means that $\emptyset_I(t)$ and $\emptyset_q(t)$ are orthogonal over a bit/symbol duration $T_s$ and each of them has a unit energy. Assume that the periodicity of the sawtooth signal is an integer multiple of the coherence time $T_c = mT_s$. In addition to this assumption, assume that the symbol duration is an integer multiple of the carrier period. By checking the orthogonality:

$$\int_{-T_s/2}^{T_s/2} \emptyset_I(t) \emptyset_q(t) \, dt$$

$$= \int_{-T_s/2}^{T_s/2} \sqrt{\frac{2}{T_s}} \cos(2\pi f_c t + K_f \int_{-\infty}^{t} \emptyset(\lambda) d\lambda) \sqrt{\frac{2}{T_s}} \sin(2\pi f_c t + K_f \int_{-\infty}^{t} \emptyset(\lambda) d\lambda), dt$$

$$= \int_{-T_s/2}^{T_s/2} \frac{1}{T_s} \cos(4\pi f_c t + 2K_f \int_{-\infty}^{t} \emptyset(\lambda) d\lambda) \, dt$$

$$= \frac{1}{T_s} \cdot \frac{\sin(4\pi f_c t + 2K_f \int_{-\infty}^{t} \emptyset(\lambda) d\lambda)}{4\pi f_c + 2K_f K t} \bigg|_{-T_s/2}^{T_s/2}$$

$$= \frac{1}{T_s} \cdot \left( \frac{\sin(2\pi f_c T_s + 2K_f \int_{-\infty}^{T_s/2} \emptyset(\lambda) d\lambda)}{4\pi f_c + K_f K T_s} - \frac{\sin(-2\pi f_c T_s + 2K_f \int_{-\infty}^{-T_s/2} \emptyset(\lambda) d\lambda)}{4\pi f_c - K_f K T_s} \right)$$

$$= \frac{1}{T_s} \cdot \left( \frac{\sin(2\pi f_c T_s)}{4\pi f_c + K_f K T_s} - \frac{\sin(-2\pi f_c T_s)}{4\pi f_c - K_f K T_s} \right) = 0 \quad (5)$$

The next check is that the orthonormal basis functions have unity energy:

$$\| \emptyset_I(t) \|$$

$$= \int_{-T_s/2}^{T_s/2} \emptyset_I^2(t) \, dt$$

$$= \int_{-T_s/2}^{T_s/2} \frac{2}{T_s} \cos^2(2\pi f_c t + K_f \int_{-\infty}^{t} \emptyset(\lambda) d\lambda)$$

$$= \int_{-T_s/2}^{T_s/2} \frac{2}{T_s} (\frac{1}{2} + \frac{1}{2} \cos(4\pi f_c t + 2K_f \int_{-\infty}^{t} \emptyset(\lambda) d\lambda)) \, dt$$

$$= \frac{2}{T_s} (\frac{t}{2} \bigg|_{-T_s/2}^{T_s/2} + \frac{1}{2} \int_{-T_s/2}^{T_s/2} \cos(4\pi f_c t + 2K_f \int_{-\infty}^{t} \emptyset(\lambda) d\lambda) \, dt)$$

$$= \frac{2}{T_s} (\frac{T_s}{2} + 0)$$

$$= 1 \quad (6)$$

An M-ary PSK constellation can be represented in terms of the in-phase and quadrature-phase basis functions. The orthonormality of the basis functions is maintained and all symbols are guaranteed for a PSK constellation where all symbols lie on a constant energy circle as shown in figure 7.

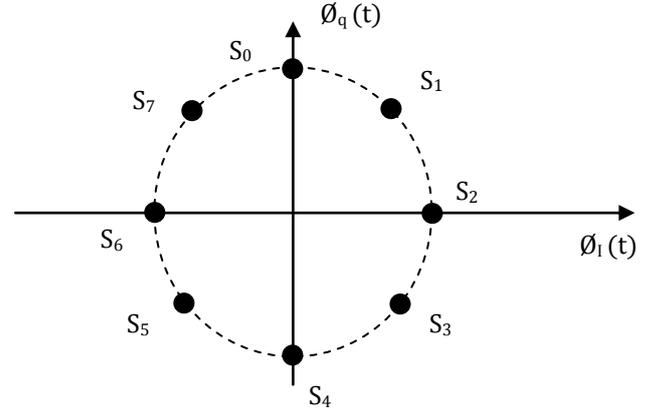

Figure 7 an example for 8-PSK constellations based on the FM sweeping basis functions

Taking a Binary PSK (BPSK) system as a case study, we have a binary constellation with two constellation points of projections $\sqrt{Es}$ and $-\sqrt{Es}$ for the bits '1' and '0' respectively, where Es is the symbol energy. The two constellation symbols are:

$$S(t) =$$

$$\begin{cases} \sqrt{\frac{2Es}{T_s}} \cos\left(2\pi f_c t + K_f \int_{-\infty}^{t} \emptyset(\lambda) d\lambda\right) \text{ for } '1', \ 0 \leq t \leq T_s \\ -\sqrt{\frac{2Es}{T_s}} \cos\left(2\pi f_c t + K_f \int_{-\infty}^{t} \emptyset(\lambda) d\lambda\right) \text{ for } '0', \ 0 \leq t \leq T_s \end{cases}$$

$$= \begin{cases} \sqrt{Es} \, \emptyset_I(t) \text{ for } '1', \ 0 \leq t \leq T_s \\ -\sqrt{Es} \, \emptyset_I(t) \text{ for } '0', \ 0 \leq t \leq T_s \end{cases} \quad (7)$$

The block diagram of the BS-2-PSK modulator is shown in figure 8. The sinusoidal carrier is FM modulated by a ramp function (saw tooth) and then used to modulate the binary polar Non-return to Zero (NRZ) symbols.

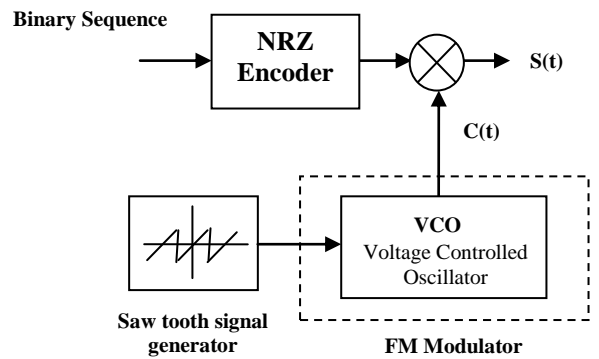

Figure 8 block diagram for a basic BS-2-PSK modulator

## III. CHANNEL ESTIMATION AND RECEIVER STRUCTURE

In this section, we start by carrying out the analytical spectral analysis for the BS-M-PSK signal, and then we present the receiver structure together with the channel estimation algorithm.

### A. The Spectrum of the FM Modulated Carrier

The spectrum of the modulated carrier is derived based on Fourier analysis. Based on equation (2), the signal is a periodic sinusoidal with a periodic time varying argument. Thus, the overall signal C (t) is periodic and can be represented based on the Fourier series representation. By obtaining the Fourier series representation of the saw tooth function, the synthesis equation is:

$$\emptyset(t) = \sum_{n=-\infty}^{\infty} F_n e^{\frac{jn2\pi t}{T_c}} \quad (8)$$

Where $F_n$ are the Fourier series coefficients. The analysis equation is then used to obtain the coefficients:

$$F_n = \frac{K}{T_c} \int_{-T_c/2}^{T_c/2} t \, e^{-\left(\frac{jn2\pi t}{T_c}\right)} dt \quad (9)$$

Applying integration by parts, the exponential Fourier series coefficients become of a sawtooth signal are:

$$F_n = \frac{jKT_c}{2\pi n} \cos(n\pi)$$

$$= \frac{jKT_c}{2\pi n} (-1)^n, \; n = ...,-2,-1,1,2,3,.... \quad (10)$$

and we can re-write the sawtooth signal as:

$$\emptyset(t) = \sum_{n=-\infty,\, n\neq 0}^{\infty} \frac{jKT_c}{2\pi n} (-1)^n e^{\frac{jn2\pi t}{T_c}} \quad (11)$$

The FM modulated carrier can be re-written as:

$$C(t) = \cos(2\pi f_c t + K_f \int_{-\infty}^{t} \sum_{n=-\infty,\, n\neq 0}^{\infty} \frac{jKT_c}{2\pi n} (-1)^n e^{\frac{jn2\pi \lambda}{T_c}} d\lambda)$$

$$= \text{RE}\{e^{j2\pi f_c t} e^{j(K_f \int_{-\infty}^{t} \sum_{n=-\infty,\, n\neq 0}^{\infty} \frac{jKT_c}{2\pi n} (-1)^n e^{\frac{jn2\pi \lambda}{T_c}} d\lambda)}\} \quad (12)$$

Recall that the Taylor series expands the exponential function as:

$$e^x = 1 + x + \frac{x^2}{2!} + \frac{x^3}{3!} + ...... + \frac{x^m}{m!} + .....$$

$$= \sum_{m=0}^{\infty} \frac{x^m}{m!} \quad (13)$$

Thus, the modulated carrier can be expressed as:

$$C(t) = \text{RE}\{e^{j2\pi f_c t} \sum_{m=0}^{\infty} \frac{(jK_f \int_{-\infty}^{t} \sum_{n=-\infty,\, n\neq 0}^{\infty} \frac{jKT_c}{2\pi n} (-1)^n e^{\frac{jn2\pi \lambda}{T_c}} d\lambda)^m}{m!}\} \quad (14)$$

Assuming that only the first two terms are significant (Narrow-band FM case), and that only two Fourier coefficients are significant for the sawtooth signal. We can simplify the modulated carrier as:

$$C(t) = \text{RE}\{e^{j2\pi f_c t}(1 + jK_f \int_{-\infty}^{t} \sum_{n=-\infty,\, n\neq 0}^{\infty} \frac{jKT_c}{2\pi n} (-1)^n e^{\frac{jn2\pi \lambda}{T_c}} d\lambda)\}$$

$$\approx \text{RE}\{e^{j2\pi f_c t}(1 + jK_f \int_{-\infty}^{t}\left(-\frac{KT_c}{j2\pi} e^{\frac{-j2\pi \lambda}{T_c}} + \frac{KT_c}{j2\pi} e^{\frac{j2\pi \lambda}{T_c}}\right) d\lambda)\}$$

$$= \text{RE}\{e^{j2\pi f_c t}(1 + j\frac{K_f K T_c}{\pi} \int_{-\infty}^{t} \sin\left(\frac{2\pi \lambda}{T_c}\right) d\lambda)\}$$

$$= \text{RE}\{e^{j2\pi f_c t}(1 - j\frac{K_f K T_c^2}{2\pi^2}(1 + \cos(\frac{2\pi t}{T_c})))\}$$

$$= \text{RE}\{(\cos(2\pi f_c t) + j\sin(2\pi f_c t))(1 - j\frac{K_f K T_c^2}{2\pi^2}(1 + \cos(\frac{2\pi t}{T_c})))\}$$

$$= \text{RE}\{(\cos(2\pi f_c t) + j\sin(2\pi f_c t) - j\frac{K_f K T_c^2}{2\pi^2} \cos(2\pi f_c t) + \frac{K_f K T_c^2}{2\pi^2} \sin(2\pi f_c t) - j\frac{K_f K T_c^2}{2\pi^2} \cos(2\pi f_c t) \cos(\frac{2\pi t}{T_c}) + \frac{K_f K T_c^2}{2\pi^2} \sin(2\pi f_c t) \cos(\frac{2\pi t}{T_c}))\}$$

$$= \cos(2\pi f_c t) + \frac{K_f K T_c^2}{2\pi^2} \sin(2\pi f_c t) + \frac{K_f K T_c^2}{2\pi^2} \sin(2\pi f_c t) \cos(\frac{2\pi t}{T_c})$$

$$= \cos(2\pi f_c t) + \frac{K_f K T_c^2}{2\pi^2} \sin(2\pi f_c t) + \frac{K_f K T_c^2}{4\pi^2}(\sin(2\pi (f_c + 1/T_c)t) - \sin(2\pi (f_c - 1/T_c)t)) \quad (15)$$

By applying Fourier transform to this expression, the spectrum of the FM modulated carrier is given by:

$$C(f) = \pi[\delta(f - f_c) + \delta(f + f_c)] + \frac{K_f K T_c^2}{j2\pi}[\delta(f - f_c) - \delta(f + f_c)] + \frac{K_f K T_c^2}{j4\pi}([\delta(f - (f_c + 1/T_c)) - \delta(f + (f_c + 1/T_c))] - [\delta(f - (f_c - 1/T_c)) - \delta(f + (f_c - 1/T_c))]) \quad (16)$$

Assume that the transmission bandwidth for the transceiver is denoted by $B/2\pi$ rad/sec. The FM modulated carrier must scan the whole range of transmission. The minimum instantaneous frequency is $\omega_c - K_f K T_c /2$ and the maximum frequency is $\omega_c + K_f K T_c /2$ and thus the next condition is imposed on the constants $K$ and $K_f$:

$$K_f K T_c = B \quad (17)$$

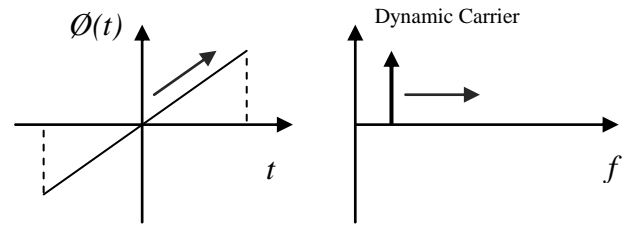

Figure 9 as the saw tooth function increases in value with time, the carrier tone (not exactly an impulse in frequency domain due to limited constant frequency duration) sweeps across higher frequency components

## B. The Receiver Architecture

The receiver of the BS-2-PSK is based on a *correlator*, a Channel estimator, an Equalizer and a slicer (decision device).

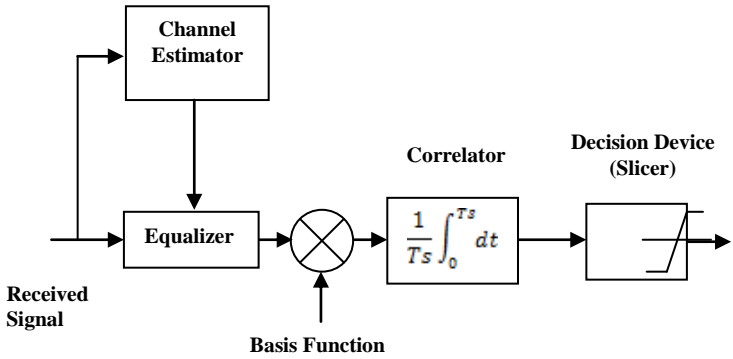

Figure 10 the receiver structure for BS-2-PSK system

The channel estimator is composed of an envelope detector circuit. A rectifier is followed by a well designed Low Pass Filter (LPF) will trace the envelope of the received FM modulated carrier. The channel estimate is then passed to a conventional equalizer that relies on inverting the channel (ZF equalizer), or any other equalization technique.

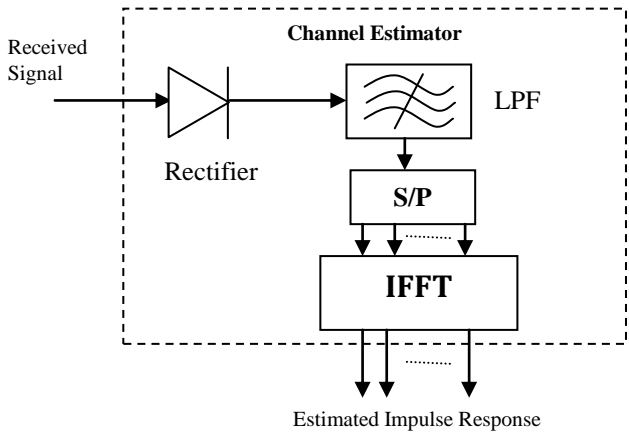

Figure 11 the channel estimation block diagram

By applying a two-tap Rayleigh fading channel with delay spread of 1 μsec and a relative -4 dB power for the indirect path, assuming a carrier frequency of 0.1 Mrad/sec and 31.25 msec coherence time, $K_f = 120$, $K = 0.75$, the estimated channel is shown in figure 12. The channel estimate is then applied to an IFFT (Inverse Fast Fourier Transform algorithm) and the channel tap gains are passed to an equalizer (ZF, MMSE, DFE,..etc) where the Inter-Symbol Interference (ISI) is removed from consecutive PSK symbols.

## IV. CONCLUSION

A new class of digital modulation is presented in this work. The modulation scheme (called Band-Sweeping M-ary PSK or BS-M-PSK) facilitates the channel estimation process at the receiver end by varying the instantaneous frequency of the carrier between the minimum and maximum frequencies in the transmission band. This is achieved by applying FM modulation to the RF carrier with a sawtooth signal that is periodic with a period that is equal to the channel coherence time. This variation makes the channel gain be proportional to the channel transfer function, and thus, the envelope of the received FM modulated carrier has the same shape of the channel transfer function. The transceiver block diagram of a BS-2-PSK system was presented and the channel estimation algorithm was shown.

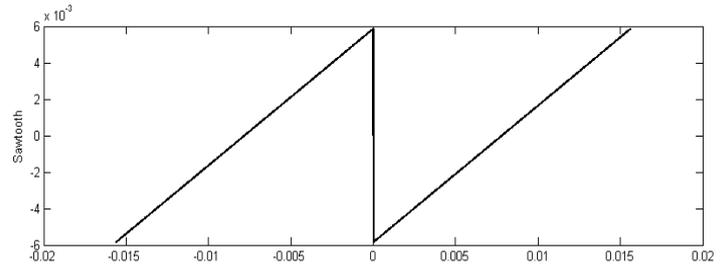

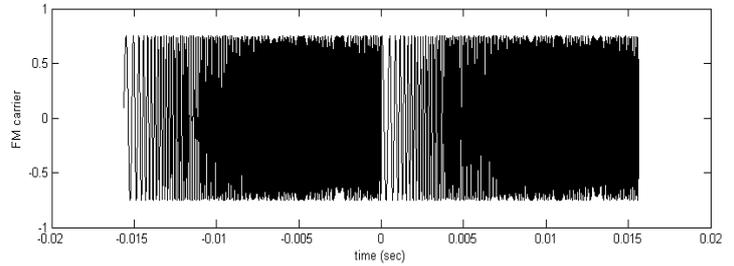

(a)

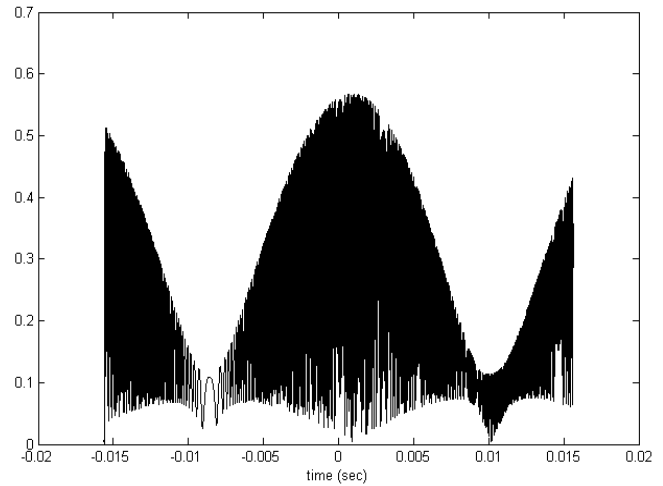

(b)

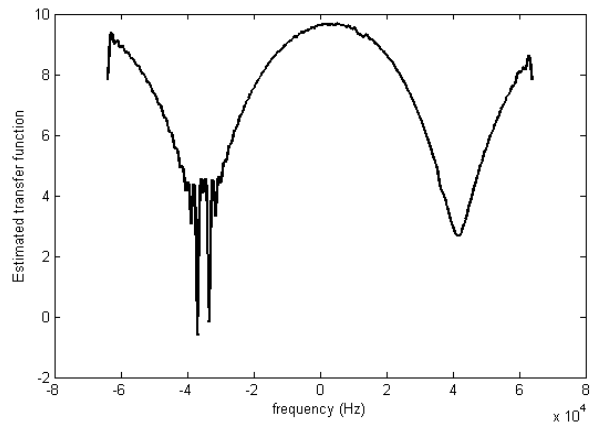

(c)

Figure 12 (a) the saw tooth signal and the FM modulated carrier (b) the received sweeping carrier (c) the estimated transfer function

## About the Author

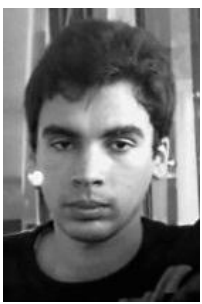

Ahmed M. Alaa received BSc from Cairo University in 2011. His research interests include: Physical layer modeling for wireless systems, Analog communications, Mobile Networks planning and optimization, Information theory and receiver design. His work is focused on developing receiver algorithms for multicarrier systems (OFDM and FBMC). He is currently a Teaching and Research Assistant at Cairo University.